\documentclass[10pt, a4paper]{article}

\usepackage{microtype}
\usepackage{inconsolata}
\usepackage{ragged2e}
\usepackage{subcaption}
\usepackage{graphicx}
\usepackage{nicematrix}
\usepackage{multirow}
\usepackage[export]{adjustbox}
\usepackage{todonotes}
\usepackage[framemethod=TikZ]{mdframed}
\usepackage{xcolor}
\usepackage{tabularx}
\usepackage{soul}
\usepackage{listings}
\usepackage{caption}
\usepackage{booktabs}

\newmdenv[%
    backgroundcolor=gray!10,
    linecolor=black,
    outerlinewidth=0.5pt,
    roundcorner=1mm,
    skipabove=\topsep,
    skipbelow=\topsep,
    font=\ttfamily\tiny,
]{promptbox}

\usepackage{lrec-coling2024}
\title{DOSA: A Dataset of Social Artifacts from Different Indian Geographical Subcultures}

\name{Agrima Seth$^{1,*}$, Sanchit Ahuja$^{2}$, Kalika Bali$^{2}$, Sunayana Sitaram$^{2}$} 

\address{$^{1}$School of Information, University of Michigan, \\
$^{2}$Microsoft Research India\\
agrima@umich.edu, \{t-sahuja, kalikab, sunayana.sitaram\}@microsoft.com \\}

\abstract{
Generative models are increasingly being used in various applications, such as text generation, commonsense reasoning, and question-answering. To be effective globally, these models must be aware of and account for local socio-cultural contexts, making it necessary to have benchmarks to evaluate the models for their cultural familiarity. Since the training data for LLMs is web-based and the Web is limited in its representation of information, it does not capture knowledge present within communities that are not on the Web. Thus, these models exacerbate the inequities, semantic misalignment, and stereotypes from the Web. There has been a growing call for community-centered participatory research methods in NLP. In this work, we respond to this call by using participatory research methods to introduce \textit{DOSA}, the first community-generated \textbf{D}ataset \textbf{o}f 615 \textbf{S}ocial \textbf{A}rtifacts, by engaging with 260 participants from 19 different Indian geographic subcultures. We use a gamified framework that relies on collective sensemaking to collect the names and descriptions of these artifacts such that the descriptions semantically align with the shared sensibilities of the individuals from those cultures. Next, we benchmark four popular LLMs and find that they show significant variation across regional sub-cultures in their ability to infer the artifacts.
\newline \Keywords{Generative AI, LLMs, social artifacts, human-centered dataset creation, participatory research, Global South, non-western dataset} }

\begin{document}

\maketitleabstract
\def\thefootnote{*}\footnotetext{Work done while at Microsoft Research India}
\section{Introduction}
\label{sec:intro}
Large Language Models (LLMs) are increasingly being integrated with various applications that have a direct social impact \citep{tamkin2021understanding}, such as chatbots for health advice \citep{jo2023understanding, cabrera2023ethical} and content moderation \citep{wang2023evaluating}. There has been an increase in the concerns about what cultural nuances they have, what world knowledge they encode, what ideologies their outputs mimic \citep{atari2023humans}, and what gender \citep{thakur2023unveiling,kotek2023gender}, race \citep{fang2023bias}, political identities \citep{motoki2023more}, and experiences are these models aware of. LLMs trained on large-scale, diverse, and filtered web data are considered by many as adept in performing multiple tasks \citep{brown2020language}. However, the Web itself is lacking and inequitable, i.e., while certain cultures and their related knowledge are represented more than others, the knowledge of many cultures is missing altogether. Various scholarships on a wide variety of tasks, such as question-answering \citep{palta2023fork}, value alignment, and fairness, have shown that these models predominantly align with the Western, Educated, Industrialized, Rich, and Democratic (WEIRD) ideologies \citep{atari2023humans}, are Anglo-centric and reproduce some harmful stereotypes and biases \citep{thakur2023unveiling,kotek2023gender,abid2021large}
\begin{figure}[ht!]
    \includegraphics[width=\columnwidth]{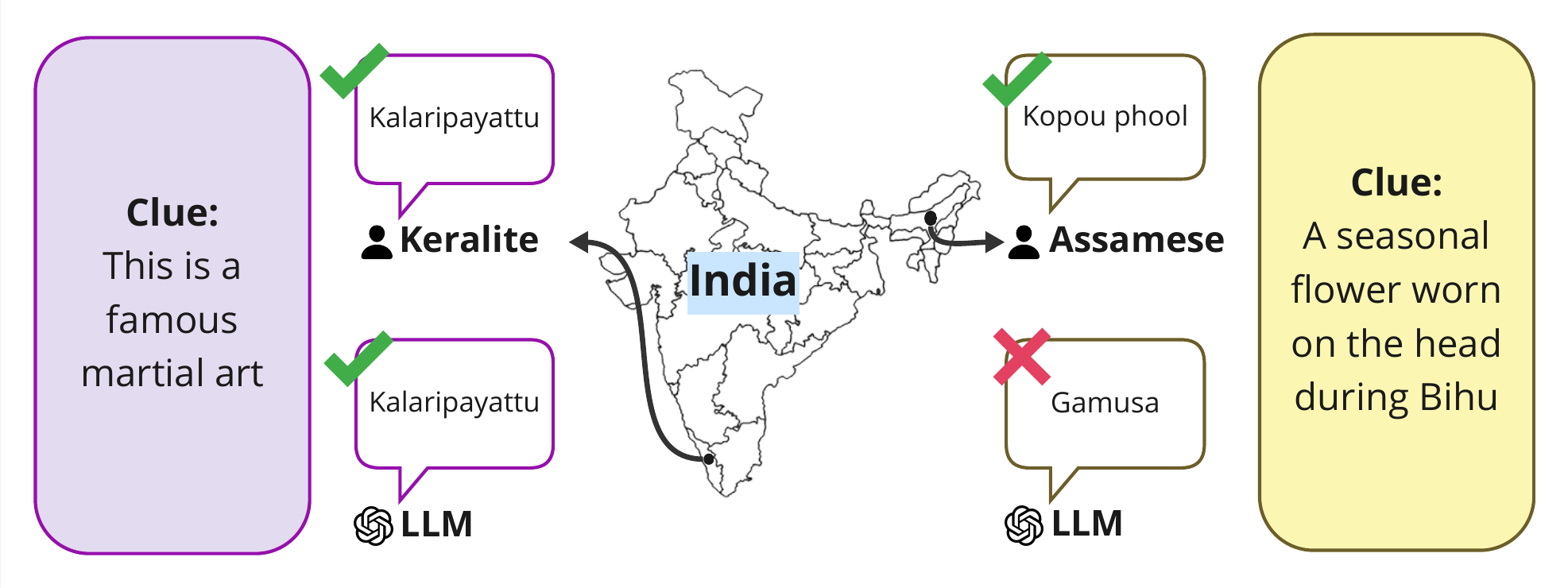}
    \caption{Are LLMs equally aware of social artifacts from different subcultures within a country? For social artifacts from different state-based subcultures of India, we prompt LLMs with unique information that differentiates a social artifact from others and evaluate their overall accuracy for each state.}
    \label{fig:enter-label}
\end{figure}
 
For LLMs to have global acceptability, we must understand what cultural knowledge and behaviors their outputs mimic. There are multiple dimensions to understanding the cultural acceptability of LLMs. One dimension that has been studied in the past to assess LLMs applicability for decision-making is cultural values like individualism and collectivism, authority and subversion \citep{hofstede2011dimensionalizing,graham2013moral,seth2023cultural}. However, (a) for tasks on content production like text generation and creative writing, (b) to avoid risks like cultural erasure by omission and propagating hegemonic views language models, and (c) to not place \textit{extra burden} of communication on members of non-Euro-centric cultures, these models need to be aware of the social artifacts and the commonplace knowledge associated with them that is present in the target society and actively use this knowledge in content production \cite{prabhakaran2022cultural} \footnote{For example, ``Dosa," a crispy, savory dish in southern states of India, might be called a crepe-like dish; however, the two are not equivalent, and the community where Dosa comes from would not use the linguistic formation ``crepe-like" to describe it. Yet the Wikipedia definition takes a Euro-centric stance to define this social artifact \url{https://en.wikipedia.org/wiki/Dosa_(food)}}. However, this dimension of cultural awareness and alignment is understudied.


Surprisingly, there has not been a systematic evaluation of whether and how the existing models encode the knowledge about artifacts considered essential and commonplace by individuals from the culture. There are many challenges in creating robust evaluation datasets for cultural understanding. First, getting data across different cultures is difficult from the Web because not all cultures find equal representation on the Web. Even when artifacts are represented on the Web, they are likely to be those that have been embraced or recognized by other mainstream societies and many-a-times are remnants or reproduce colonial knowledge \citep{mamadouh2020writing}, which further diminishes the voices of the community members and might propagate stereotypes \cite{qadri2023ai}. The second challenge is accessing knowledge sources and people from whom cultural data can be collected and determining how to meaningfully involve individuals from different cultures in dataset creation and multi-cultural research. While past work in multilingual studies created parallel datasets using translation as a strategy, some culture-specific concepts and objects often do not have equivalents in other cultures and hence either do not have a linguistic equivalent or the semantics do not correlate with the sensibilities of community members\cite{hershcovich2022challenges}. Thus, the creation of culture-based datasets and subsequent evaluations of LLMs is a challenging task.

In this work, we use bottom-up, community-centered participatory research methods in a non-Western context and engage with community members to introduce a dataset of 615 social artifacts' names and descriptions across 19 regional subcultures of India. We use surveys and implement a gamified framework to create a dataset of social artifacts and use this dataset to benchmark the cultural familiarity of the four most widely used and recent LLMs - GPT4\cite{openai2023gpt4}, LlAMA2\cite{touvron2023llama}, PALM 2\cite{anil2023palm}, and FALCON \cite{falcon40b}. In particular, this work focuses on the diverse culture of India - a country in the Global South.\\

\noindent \textbf{Contributions:} Past work positions that language technologies can benefit from integrating community intelligence via participatory research \citep{diddee2022six}. However, \textbf{\textit{``how"}} one designs participatory frameworks for effective involvement of users is a non-trivial problem. In this paper, we respond to the calls for more community-centered research and show how participatory research methods can effectively create datasets. Second, this is the first paper that explores cultures at a geographical level within a country - India and presents a social artifact dataset to expand the field's understanding of the cultural familiarity of LLMs. Next, we benchmark four widely used LLMs (both open and closed sources) and find a significant inter-LLM variation in their familiarity with the social artifacts. Third, we discuss the obstacles and learnings derived from engaging with participatory research to create evaluation datasets. Thus, our work offers an example of how technology evaluation can benefit from engaging community members using participatory research. 


Culture is a complex societal-level concept, and it can be defined by multiple factors: location, sexuality, race, nationality, language, religious beliefs, ethnicity, etc. Past work has shown the significance of geographic boundaries in determining cultural identity, such as the World Value Survey \citep{wvs}. This study focuses on studying the cultural identities based on India's geographic states. India has 28 geographic states, each with different languages, food, and customs, many of which also do not find appropriate representation on the Web. This is the first attempt to use participatory research to collect social artifacts that are commonplace and perceived as important by the members of the respective communities. While this is not a complete dataset of all important social artifacts, future work could draw from this paper to design participatory research-based methods to scale the dataset to more subcultures.

\section{Related Work}
\label{sec:rw}

\subsection{Cultural Awareness of Language Models}
\label{sec:awareness}
Past work on LLM evaluations has focused on understanding the personality, values, and ethics encoded in these models. These works have probed these language models using established psychometric and cultural instruments like IPIP-NEO and the Revised NEO Personality Inventory \citep{safdari2023personality}, Moral Direction framework, Moral Foundation Theory, Hofstede's cultural dimensions, and Schwartz's cultural value \citep{yao2023instructions,schramowski2022large,arora2022probing,hammerl2022speaking,fischer2023does}. Some past works have also leveraged questions on ethical dilemmas \citep{tanmay2023exploring} to elicit what human values LLMs mimic in their output. Prior work in fairness has looked at the stereotypes and biases encoded in these models towards specific communities \cite{thakur2023unveiling,kotek2023gender,abid2021large}. In contrast to these works, which look at what ideologies LLMs have learned and the biases in the training data for LLMs, our work focuses on creating a community-centered dataset of artifacts across different regional sub-cultures to evaluate LLMs' knowledge and its alignment with the shared body of knowledge that is held in common and considered important by the members of the respective cultural communities. 


The most closely related work to ours is \citet{acharya2020towards}, which used surveys to get information on four specific rituals from MTurkers in the USA and India, and \citet{nguyen2023extracting}, which scrapped Wikipedia to create a knowledge graph of commonsense knowledge of geography, religion, occupation and integrated it with LLMs. However, to the best of our knowledge, ours is the first work that leverages bottom-up participatory research to build a dataset of social artifacts that evaluates the alignment of commonplace knowledge of community members and LLMs and does not use pre-defined categories to restrict its participants.

\subsection{Participatory Research for Dataset Creation}
\label{sec:participatory}
Participatory research argues that individuals who are affected by technology should be involved in designing and evaluating it. Human-computer interaction (HCI) as a field has extensively used surveys, focus groups, and interviews, but surprisingly, the use of participatory methods in NLP is largely lacking \citep{diddee2022six}. Prior work in knowledge elicitation has shown success in using ``games with a purpose (GWAP)" for collecting commonplace knowledge and verifying concepts and their relations \citep{balayn2022ready,von2006verbosity}. An essential aspect of social artifacts is the shared knowledge and understanding of the artifact's significant aspects, use, and unique and differentiating characteristics (Refer Fig \ref{fig:enter-label}) \citep{stephenson2023culture}. Furthermore, leveraging GWAP allows us to collect more implicit knowledge based on concepts and mental models that would be otherwise hard to codify in a formal written language, usually found in datasets based on scrapping data from the Web. Thus, in this work, we use two methods of participatory research - Surveys and GWAP. Drawing inspiration from prior GWAP, we formulate a modified version of the classic Taboo game (Refer Section \ref{sec:game}) to elicit cultural artifacts and knowledge. While the previously proposed games \citep{balayn2022ready} restrict the players to specific templates, in our work, we relax this requirement by allowing them to formulate clues in natural speech.

\begin{figure*}
    \centering
    \begin{subfigure}{\textwidth}
\includegraphics[width=\textwidth, height = 0.3\textwidth]{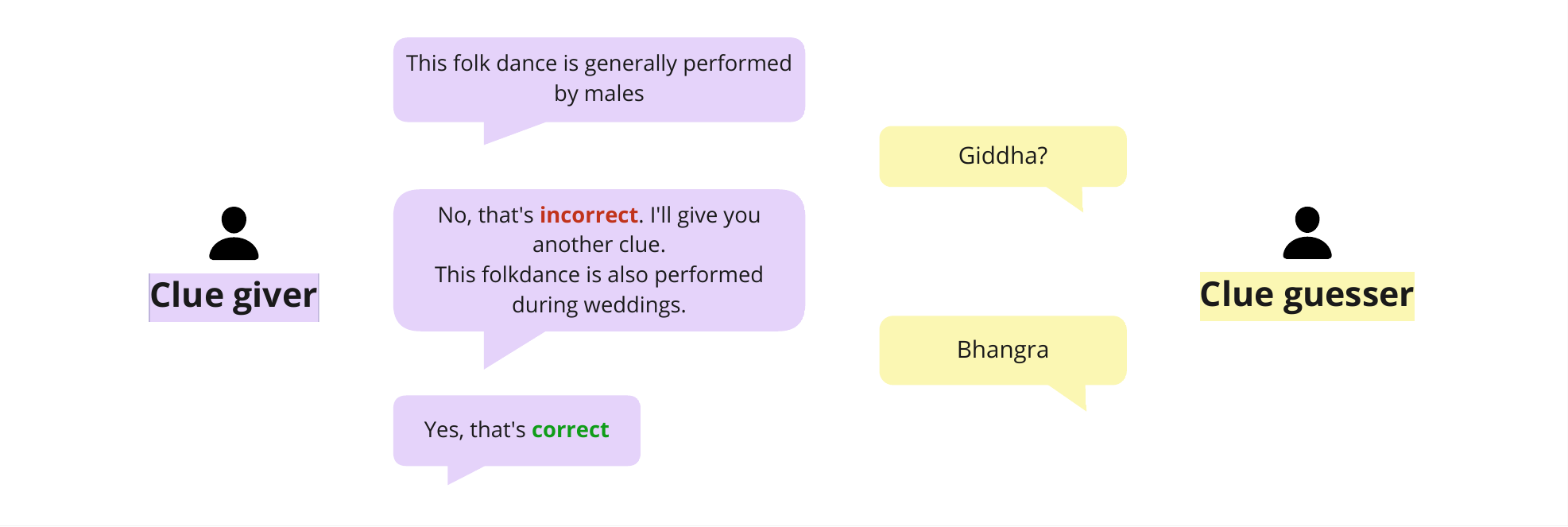}
        \caption{Illustration of the game mechanics and the knowledge elicited from the participants (from Punjab).}
    \label{fig:human_convo}
    \end{subfigure}
    \begin{subfigure}{\textwidth}
\includegraphics[width=\textwidth, height = 0.3\textwidth]{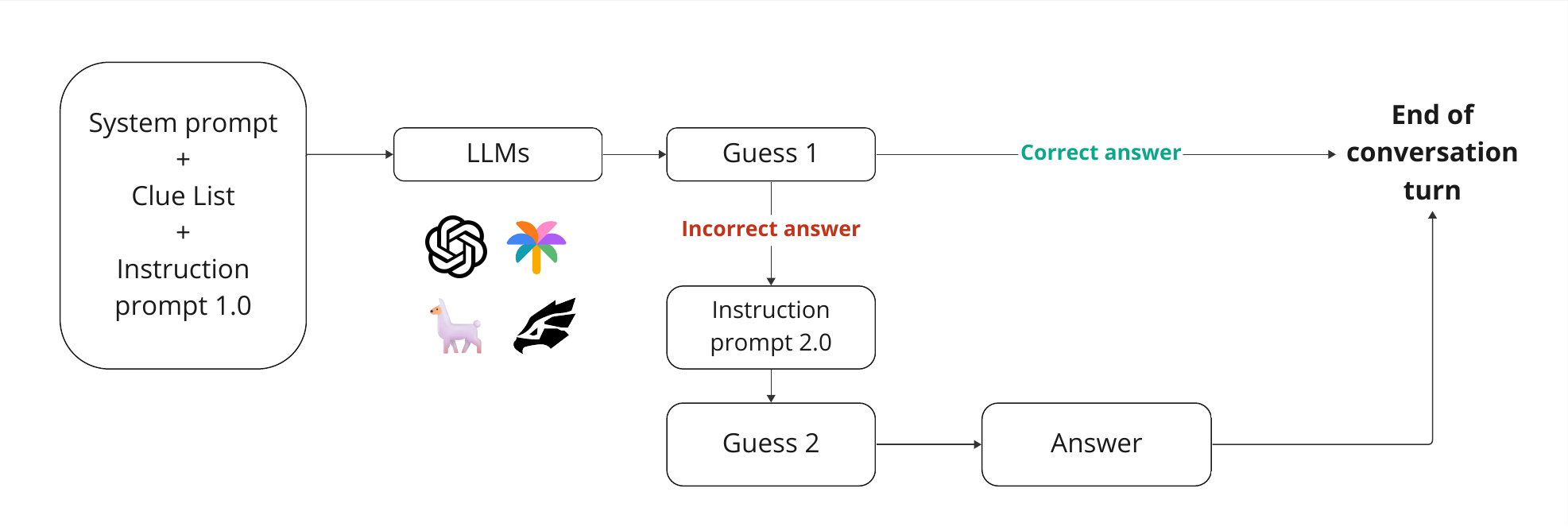}
        \caption{Illustration of how LLMs were prompted for benchmarking.}
    \label{fig:llm_convo}    
    \end{subfigure}
    \caption{Illustrations highlighting the game mechanics for knowledge elicitation and LLM benchmarking for cultural familiarity}
\end{figure*}

\section{Methodology for the Dataset Creation}
\label{sec:data}

To record information grounded in the community's shared knowledge that differs from the typical article-like written knowledge found in datasets crawled from the Web, we collect the data of social artifacts by combining two participatory research methods - \textit{Survey and Games with a Purpose (GWAP)}. First, we administer the survey to participants across the 19 Indian States, asking them to self-identify their cultural identity and name social artifacts considered important in that subculture. Next, we use these artifacts to design a GWAP and recruit participants from these 19 Indian States to provide information about the artifacts. The game also primes the participants, and when asked to volunteer to share more artifacts and their descriptions, post-game helped us expand the dataset to 615 artifacts across 19 states.

\subsection{Target Cultures}
India has 28 states and 8 Union Territories, each with diverse food, handicrafts, dance forms, festivals, rituals, and practices. There is a myriad of operationalizations for the construct of `culture,' and combining each is beyond the scope of this work. This study focuses on studying the cultural identities based on India's geographic states because the states in India were demarcated on the lines of linguistic and ethnic identities (States Reorganisation Act, 1956)\footnote{https://pwonlyias.com/upsc-notes/state-reorganisation-act/}, which have a strong correlation to a shared sense of culture in individuals from similar geographic regions \citep{singh2009unfolding}. Our study relied on the population data from the World Value Survey (WVS) Wave 6 \cite{wvs} in 2014. The survey was conducted in 18 States of India, stating that 95\% of India's population resided in those 18 states. However, the state of Telangana is a new state that was formed after the last WVS was conducted in India. Hence, we decided to collect data from 19 states --- the 18 states mentioned in the WVS and the state of Telangana.

\subsection{Pilot Study}
For the pilot experiment, we recruited 14 participants. It was an in-house pilot with employees at our research lab as the participants. Participants were asked to self-identify the state that best represents their cultural identity. In the pilot study, we had participants from 6 geographic cultures. First, these participants took the survey questionnaire and then engaged with the gamified framework. The participants were compensated for their time. Using the learnings from our pilot, we iterated over our survey questions and game design and finalized our methodology to gather data at a larger scale.

\subsection{Survey Questionnaire}
\label{sec:survey}
We used Karya Inc \footnote{\url{https://www.karya.in/}}, an ethical data company that engages economically disadvantaged Indians in digital work, to asynchronously administer five surveys across the 19 states in India, making it a total of 95 surveys. Due to operational issues, data from Madhya Pradesh could not be gathered. Each state has its official language. For example, Kannada is the official language of Karnataka, while Hindi and Urdu are the official languages of Uttar Pradesh. Past work has shown that users in India increasingly use the Roman script for online communication \citep{ahmed2011challenges,gupta2012mining}. Hence, we administered the survey in English and restricted the survey to participants with at least 12 grades of education and a working or advanced level of fluency in English. The answers for social artifacts were Romanized, i.e., written in Latin script. The compensation was decided according to the cost of living in India, and each participant was compensated with an Amazon gift voucher of INR 500 \footnote{\url{https://en.wikipedia.org/wiki/List_of_countries_by_minimum_wage}}(See Appendix \ref{sec:survey_app} for the complete survey).

As discussed earlier, cultural identity is a complex concept. To increase the clarity of instructions, provide more context to the question, and give examples of social artifacts, we added audio instructions providing more details on what cultural identity and social artifacts meant. In the survey, the survey takers were asked to self-identify the state in India that best represents their cultural identity and three states that they believe are culturally similar to theirs. Next, the participants were asked to list five social artifacts that they believe are important to their cultural identity and would be known to a reasonable number of people who share a similar geographic cultural identity. We rejected surveys where the survey takers marked the same states as the most and least similar, provided the same answers with other participants verbatim, or reiterated the same artifacts given in the instructions as examples. To help make up for the rejected surveys, the survey was re-administered. Overall, we collected 267 artifacts across all 18 states.

\subsubsection{Data Cleaning and Processing}
The survey design lends most artifacts to be transliterated, and the open-ended questions on social artifacts allow the same artifacts to be listed in varying ways. Thus, we manually reviewed the responses. At the end of this phase, the artifacts collected from the survey cover categories like names of local food cuisines, landmarks, rituals, textiles and handicrafts, dance and music forms, and literary or important political figures (Refer Table \ref{tab:artifacts_table} for examples of the artifacts). The details are in Appendix \ref{sec:clean}.

\subsection{Knowledge Extraction}
\label{sec:extract}
As members of a shared cultural community, we develop shared concepts and understandings and use them when communicating with other members of the same community. These shared concepts and understandings manifest in the language we use for communicating about these artifacts, which may vary from the more formal written information found in traditional data sources. For example, instead of referencing a famous landmark (Kashi Vishwanath Temple), we might refer to a well-known movie (Don) to communicate the place (Benaras) we are talking about. Through this game, we aim to collect knowledge about these artifacts such that it is grounded in the shared concepts and understandings of the community.

\subsubsection{Recruitment and Onboarding for the Game}
We recruited a new set of participants for this phase by broadcasting email calls to 2 educational institutes in India and reaching out to friends of friends (including parents). These participants were also asked to self-identify the state in India that best represents their cultural identity. Based on their responses, the participants were paired with another participant from the same culture. We recruited six participants from each of the 18 states to participate in the game. Each participant was compensated with an Amazon gift voucher of INR 500, and the game duration ranged between 1 hour - 90 mins. The list of artifacts obtained from the Survey (Section \ref{sec:survey}) for the corresponding state was used to conduct the game.

\subsubsection{Game Mechanics}
\label{sec:game}
\paragraph{Initialization:} At the start of the game, we shuffle the artifacts obtained for that state from the survey. Then, both participants are given a mutually exclusive list of a near-equal number of artifacts.

\paragraph{Playing the game:} Each player had two roles: (a) the one who gives the clues for the artifact - the CLUE GIVER and (b) the one who tries to identify what artifact is being talked about - the GUESSER. In each turn, the players alternate their roles. For example, let's call the players A and B. If, in the first turn, player A is the CLUE GIVER, then player B is the GUESSER, and in the next turn, player B becomes the CLUE GIVER, and player A becomes the GUESSER (Figure \ref{fig:human_convo}). The main goal for each player is to use the clues the opponent gave to guess the name of their artifact. To help elicit the most important, unique, and differentiating information about the artifact, the rules of the game were:

\noindent \textbf{1.} The CLUE GIVER could give a maximum of 5 clues in the form of a simple sentence.\\
\noindent \textbf{2.} The CLUE GIVER could not use words synonymous with the artifact.\\
\noindent \textbf{3.} The GUESSER had only two chances to make a guess.\\
\noindent \textbf{4.} The GUESSER could not ask any clarification questions. \\
During the game, the researchers primarily served as observers and ensured enforcement of the above-stated rules.
\vspace{-1em}
\paragraph{Clue formulation:} While we did not restrict the clues to be given in a templatized form, we did ask the CLUE GIVER to highlight the information that (a) most people with the shared geographic culture would be aware of or agree with, and (b) can be considered the most defining and distinctive to it. Not being restricted by a templatized format for clue formation allowed the players to generate a very rich dataset of both positive or generative knowledge (i.e., what the artifact is) and negative or discriminative knowledge (i.e., what the artifact is not).

\subsection{Post-processing of the Artifact Descriptions}
Since we recruited six participants from each culture, each artifact was described by three participants. After each game, we transcribed the clues given by the CLUE GIVER.

Across the games for a State, we observed that the clues given to identify the artifact were majorly the same, with minor differences in the wordings. After the CLUE GIVER finished giving clues and the GUESSER made their final guess, we also asked the GUESSER to rate the accuracy and quality of the clues. The perceived quality from the GUESSER and the saturation of the information in the clues over the multiple games allow us to claim reasonable validity and comprehensiveness about the artifacts' descriptions. 

\subsection{Expanding the Artifact Dataset}
To expand the list of social artifacts after the game, we asked the participants, ``What other social artifacts can be added to the list? And what would be the best way to describe them if they were a part of the list used in the game?" Since this list was expanded by conversing simultaneously with two individuals from the same culture, there was always an implicit quality check because another participant verified the artifact's name and description. Over the multiple rounds of the game, we observed that the artifacts mentioned by the participants were majorly the same. We refer to these new artifacts and their descriptions as ``expanded dataset". We will release both the original and expanded dataset of artifacts for the research community to use.

\subsection{Dataset Characteristics}
\paragraph{Participant diversity:} This study was conducted by actively recruiting individuals from 18 different Indian states. To maximize the diversity within the regions, we used Karya's platform to distribute the survey, allowing us to reach participants from lower socio-economic backgrounds in urban and rural settings. To maximize the diversity of the participants for the game, we recruited participants who are (a) attending public universities and (b) local volunteers from various NGOs. Indian public universities have a mandate for affirmative action under which they provide ``reservations" based on social and caste categories. This ensures a high diversity of participants based on gender, socio-economic status, and caste. Similarly, engaging with NGO volunteers, especially in States with a significant population of Tribes, ensures the diversity of participants.

\noindent \textbf{Diversity of Artifacts across different Identities:}
Since we used a free-form text-based survey to collect the names of artifacts and expanded them using a semi-structured interview, the artifacts collected followed a bottom-up approach, unlike prior works that relied on a more top-down approach. Hence, the DOSA dataset generated from community-centered participatory research was not limited to some pre-defined categories but encompassed a broad range of categories. Next, the diversity of participants in the survey and the game helps us ensure that the artifacts are representative of more than one community within the geographic regions.  

\begin{table}[]
\footnotesize
\begin{tabular}{@{}lccc@{}}
\toprule
\multicolumn{1}{l}{\textbf{States}} &
  \multicolumn{1}{l}{\textbf{\begin{tabular}[c]{@{}l@{}}Original\\ Artifacts\end{tabular}}} &
  \multicolumn{1}{l}{\textbf{\begin{tabular}[c]{@{}l@{}}Expanded\\ Artifacts\end{tabular}}} &
  \multicolumn{1}{l}{\textbf{Total}} \\ \midrule
Andhra Pradesh & 14 & 13 & 27  \\
Assam          & 11 & 56 & 67  \\
Bihar          & 12 & 6  & 18  \\
Chhattisgarh    & 9  & 10 & 19  \\
Delhi          & 10 & 0  & 10  \\
Gujarat        & 18 & 15 & 33  \\
Haryana        & 12 & 17 & 29  \\
Jharkhand      & 21 & 12 & 33  \\
Karnataka      & 19 & 16 & 35  \\
Kerala         & 16 & 13 & 29  \\
Maharashtra    & 16 & 18 & 34  \\
Odisha         & 12 & 32 & 44  \\
Punjab         & 20 & 25 & 45  \\
Rajasthan      & 11 & 6  & 17  \\
Tamil Nadu     & 20 & 29 & 49  \\
Telangana      & 13 & 28 & 41  \\
Uttar Pradesh  & 13 & 34 & 47  \\
West Bengal    & 20 & 18 & 38  \\ \midrule
\textbf{Total} &    &    & \textbf{615} \\ \bottomrule
\end{tabular}
\caption{Statistics regarding the number of original (from survey) and expanded (post-in-person game) artifacts available in DOSA.} 

\label{tab:artifacts}

\end{table}
\section{Benchmarking LLMs Cultural Familiarity}
\label{sec:benchmark}
We investigate whether the underlying data that large language models were trained on gives them enough context to be familiar with the social artifacts in the DOSA dataset. To make it comparable to the game (Section \ref{sec:extract}) that humans played, we prompt the models with descriptions of the artifacts from Section \ref{sec:extract} and measure their GUESSING accuracy. In this section, we summarize our experimental setup, models, and the evaluation strategy used in this work to benchmark four LLMs for their cultural familiarity.

\subsection{Experimental Setup}
We chose a balanced mix of ``popular" open source (Llama-2 \cite{touvron2023llama}, Falcon \cite{falcon40b}) and closed source (GPT-4 \cite{openai2023gpt4} and Palm 2 \cite{anil2023palm}) models to benchmark their cultural familiarity. We chose the best variants of these models that could be supported by the available compute resources, i.e., a single A100 GPU machine. We use the chat and instruct variant of Llama-2 with 13 billion parameters \cite{touvron2023llama} and Falcon with 7 billion parameters \cite{falcon40b} as the open-source models. We build an interface using Langchain \cite{Chase_LangChain_2022} to interact with these models, and to ensure reproducibility, we keep the model temperature at 0. Llama-2 and Falcon were loaded on an A100 GPU and inferred using 16-bit quantization. We used langchain to simulate a chatbot wherein we give our models the initial \hyperlink{prompt}{\textbf{System Prompt}} with the clue list and the \hyperlink{prompt}{\textbf{Instruction Prompt 1.0}}. If the FIRST GUESS is correct, the conversation turn ends, and we move on to the next artifact. If the answer is incorrect, \hyperlink{prompt}{\textbf{Instruction Prompt 2.0}} mentioning the incorrectness is given, and we ask for a SECOND GUESS (Figure \ref{fig:llm_convo}). The prompt templates are constant across models except for the special tags (from \cite{touvron2023llama} and the model documentation \cite{falcon40b}) appended to the Llama 2 and Falcon models for them to work as intended. The prompts were designed to ensure they are as close to the game conducted with the human participants to collect information about the social artifacts and ensure that the LLM has enough context to predict them.

\begin{figure}[!htbp]
    \centering
    \begin{promptbox}
    \justify
\textbf{System Prompt:} You are an agent who is well-versed in the cultures of the world. You are playing a game of taboo with another agent who is also well-versed with the cultures of the world. You can only make two guesses to identify this social artifact correctly, and you cannot ask any clarification questions. Social artifacts are objects that help us connect and stay associated with the culture. These objects are known and have significance to most people who consider themselves as a part of that culture and serve as a way of identifying themselves with the culture and the people in that culture. Your clues are: \{CLUELIST\} \\

\noindent \textbf{Instruction Prompt 1.0:} Name the object based on the above clues from \{STATE\}. I do not need to know your reasoning behind the answer. Just tell me the answer and nothing else. If you do not know the answer, say that you do not know the answer. Format your answer in the form of ANSWER: your\_answer\_here. \\

\noindent \textbf{Instruction Prompt 2.0:} Your first guess is not correct. While making your second guess, please stick to the format as ANSWER: your\_answer\_here.
    \end{promptbox}
    \caption{The instructions used for prompting the LLMs.}
    \label{fig:prompt}
\end{figure}
\vspace{-0.7cm}

\begin{figure*}
    \centering
    \begin{subfigure}{\columnwidth}
    \includegraphics[width=\columnwidth, frame]{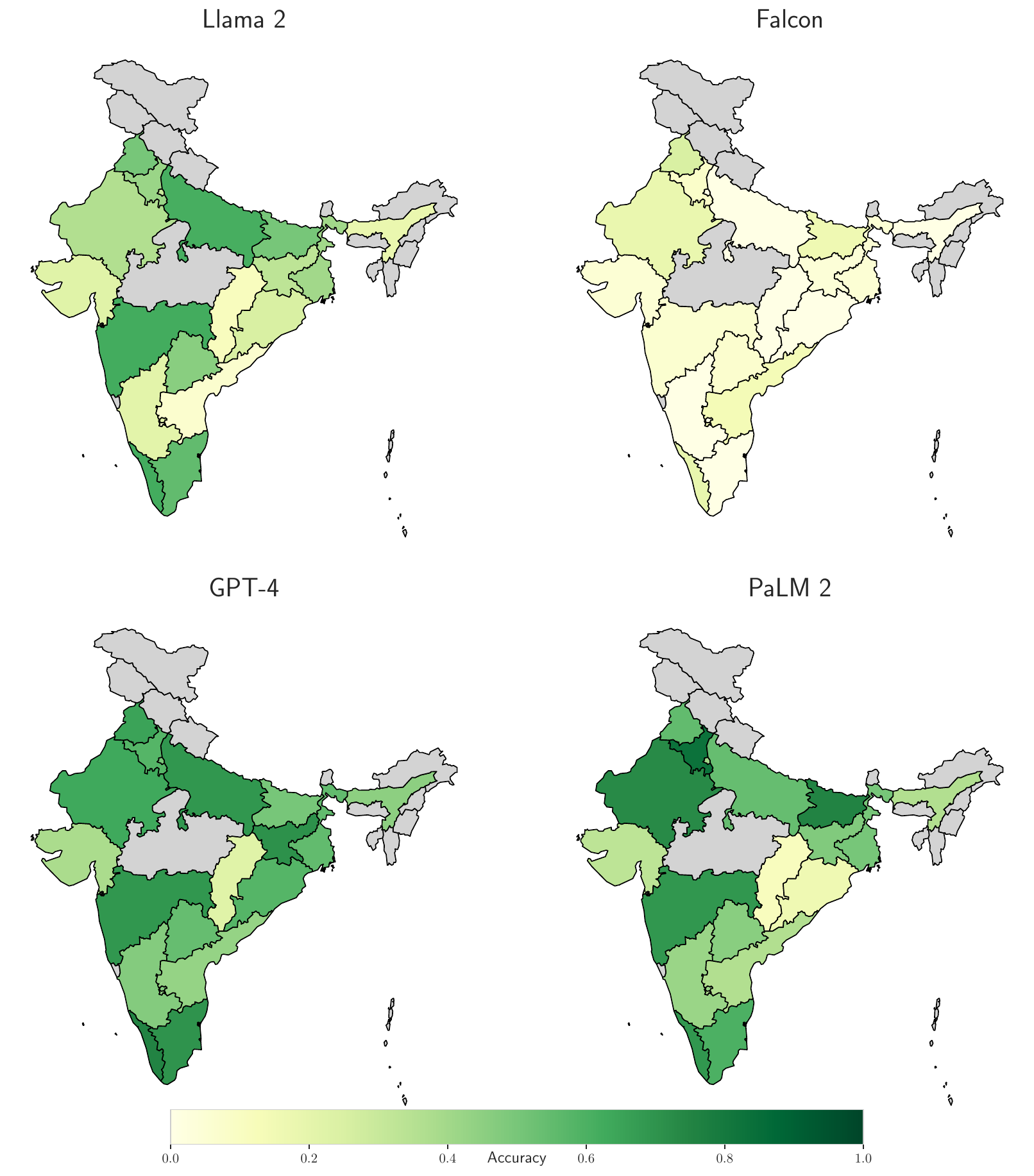}
        \caption{\textbf{Original Artifacts}}
        \label{fig:heatmap}
    \end{subfigure} \hfill
    \begin{subfigure}{\columnwidth}
    \includegraphics[width=\columnwidth, frame]{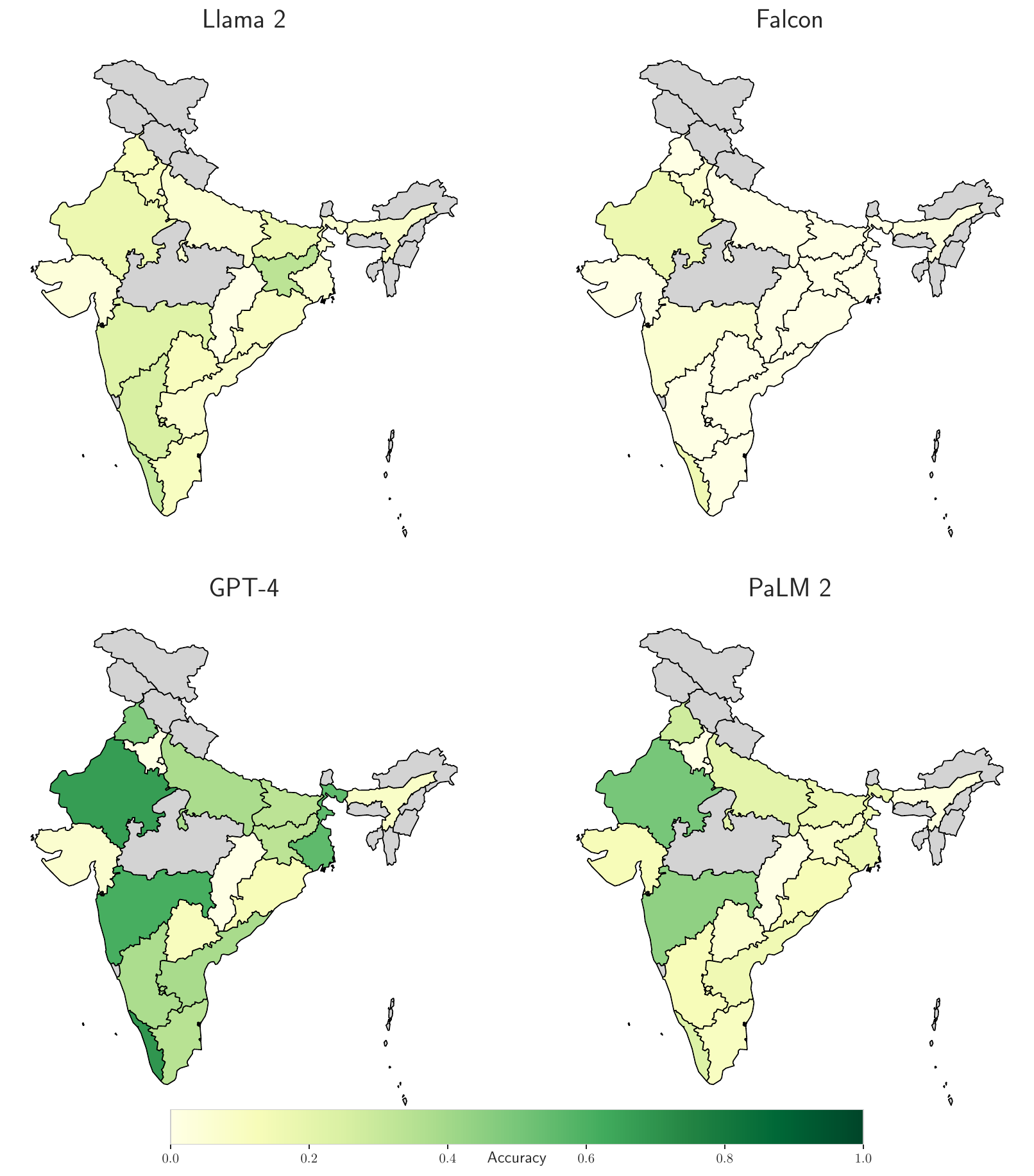}
        \caption{\textbf{Expanded Artifacts}}
        \label{fig:expanded_artifacts_heatmap}
    \end{subfigure}
    \caption{Overall accuracy of different models across 18 States in India on both the Original Artifacts (Fig \ref{fig:heatmap}) and Expanded Artifacts (Fig \ref{fig:expanded_artifacts_heatmap}). Due to operational constraints, Madhya Pradesh (denoted in grey) was excluded from data collection.}
    \label{fig:heatmap_all}
\end{figure*}

\subsection{Evaluation Setup}
\label{sec:eval}
We use accuracy as the primary evaluation metric for assessing the LLMs' cultural familiarity. We report accuracy at three levels - accuracy@GUESS1, accuracy@GUESS2, and overall accuracy. Since LLMs may sometimes produce transliterations that differ from the actual ground truth to ensure the validity of measurement to ascertain the correctness of the guess, we manually matched both the first and second guesses against the ground truth.

\noindent \textbf{1. accuracy@GUESS1}: To assess how often the model accurately predicts the artifact on the first attempt, it is calculated by dividing the number of correct FIRST GUESS predictions by the total number of social artifacts.\\
\noindent \textbf{2. accuracy@GUESS2}: This is calculated by dividing the number of correct SECOND GUESS predictions by the total number of unguessed artifacts (i.e., artifacts not predicted correctly at GUESS1).\\
\noindent \textbf{3. Overall Accuracy}: To quantify how often the model correctly predicts social artifacts (both the first and second guess, if applicable). It is calculated by dividing the number of correct predictions by the number of social artifacts.

\section{Results} 
First, the LLMs were evaluated for cultural familiarity with the artifacts collected from the survey (Section \ref{sec:survey}). Next, we evaluated their cultural familiarity with the artifacts from the expanded dataset (Refer Table \ref{tab:overall_expanded_artifacts_appendix_table}). We find that apart from the well-studied Anglo-centrism of LLMs, they significantly varied in their familiarity with regional subcultures in India, as seen in Figure \ref{fig:heatmap_all}. GPT-4 and Palm 2 perform significantly better than their open-source counterparts - Llama 2 and Falcon, with Falcon barely making any correct guesses. While GPT-4 has performed better overall, Palm 2 performs better for Bihar, Haryana and Rajasthan. We also find that across all cases, the SECOND GUESS does not lead to an increase in accuracy, implying that even when the feedback is given to the models that they are wrong, they are unable to correct themselves and get to the right artifact - an effect more pronounced for the open-source model Falcon.(Refer table \ref{tab:results_appendix} for accuracy@GUESS1 and accuracy@GUESS2)
Further, we also see that these models do not perform equally on data from each state; for example, although GPT-4 is the best-performing of all the four, it does not work at equal accuracy across all states Fig \ref{fig:heatmap}. For instance, it still cannot identify half of the artifacts from states that rank higher on the Multidimensional Poverty Index, like Assam and Chhattisgarh \citep{niti}. We then evaluated the LLMs on the expanded artifacts and found a sharp decrease in the models' performance (Refer Fig \ref{fig:expanded_artifacts_heatmap}). One hypothesis is that the artifacts we collected after the game were much more nuanced than the original artifact list from the survey. Since we manually verified the outputs of LLMs, we discovered that they misclassify culturally similar artifacts and align more towards more ``popular" artifacts. For instance, Karwa Chauth and Teej festivals involve women fasting for their husbands' long lives. However, Karwa Chauth is specific to Punjabi culture and has been widely represented in popular culture through Bollywood movies, while Teej is predominantly celebrated in Uttar Pradesh and Rajasthan. This misclassification persists even when utilizing prompt engineering techniques, which include providing explicit location information (Fig \ref{fig:prompt}) and festival-specific clues from the clue list.

\begin{table}[]
\footnotesize
\setlength{\tabcolsep}{3pt}
\begin{tabular}{@{}lcccc@{}}
\toprule
\textbf{States} & \multicolumn{4}{c}{\textbf{Models}}         \\ \midrule
 & \multicolumn{2}{c}{\textit{Open Source}}                                   & \multicolumn{2}{c}{\textit{Commercial}}                                  \\
 & \multicolumn{1}{l}{\textbf{Llama 2}} & \multicolumn{1}{l}{\textbf{Falcon}} & \multicolumn{1}{l}{\textbf{GPT-4}} & \multicolumn{1}{l}{\textbf{Palm 2}} \\ \cmidrule(l){2-5} 
Andhra Pradesh  & 0.07 & 0.14 & \textbf{0.43} & 0.36          \\
Assam           & 0.18 & 0    & \textbf{0.45} & 0.36          \\
Bihar           & 0.50 & 0.16 & 0.50          & \textbf{0.75} \\
Chhattisgarh     & 0.11 & 0    & \textbf{0.22} & 0.11          \\
Delhi           & 0.30 & 0    & \textbf{0.50}  & 0.40          \\
Gujarat         & 0.22 & 0.05 & \textbf{0.38} & 0.33          \\
Haryana         & 0.42    & 0.08    & 0.58             & \textbf{0.83}             \\
Jharkhand       & 0.33 & 0.04 & \textbf{0.71} & 0.57          \\
Karnataka       & 0.21 & 0    & \textbf{0.47} & 0.42          \\
Kerala          & 0.62 & 0.19 & \textbf{0.75} & 0.69          \\
Maharashtra     & 0.62 & 0.06 & \textbf{0.69} & \textbf{0.69} \\
Odisha          & 0.25 & 0    & \textbf{0.58} & 0.16          \\
Punjab          & 0.50 & 0.25 & \textbf{0.65} & 0.55          \\
Rajasthan          & 0.36 & 0.18 & 0.63 & \textbf{0.73}          \\
Tamil Nadu      & 0.55 & 0    & \textbf{0.70} & 0.60          \\
Telangana       & 0.46 & 0.07 & \textbf{0.54} & 0.46          \\
Uttar Pradesh       & 0.61 & 0 & \textbf{0.69} & 0.54         \\
West Bengal     & 0.40 & 0.05 & \textbf{0.55} & 0.50          \\ \bottomrule
\end{tabular}
\caption{The table denotes the overall accuracy of the LLMs on the \textbf{Original Artifacts} in DOSA.}
\label{tab:results}
\end{table}

\section{Discussion \& Future Work}
With the increase in the use of LLMs for various tasks, there has been an increase in evaluating LLMs for values, knowledge, and biases encoded in them. However, each nation itself is culturally diverse. Our work focused on assessing the LLMs at the geographic subcultural level in India. Past work shows that most models treat users from the same Euro-centric lens and assume knowledge of the Global North as the default, resulting in less representative outputs. The lack of representativeness is usually attributed to the lack of training data from ``certain cultures or regions." In this work, we created a novel knowledge dataset of social artifacts using participatory research and analyzed the cultural familiarity of the four most well-known and widely used LLMs. We find that LLMs have variance in their familiarity, but they are not entirely unaware of these artifacts. This aligns with past work that shows that LLMs prefer American cultural values in chat settings \cite {cao2023assessing}.  In multicultural contexts, chatbots violating social norms can lead to communication breakdown \citep{jurgens2023your}. Hence, it's crucial to evaluate why LLMs aren't representing their awareness of social artifacts in their outputs.

\noindent \textbf{Learnings:} The conversations and participants' feedback were precious sources of information. All participants agreed that the artifacts in the survey response were important and well-known to the community. However, they also mentioned that many of these artifacts could be considered ``popular" in India. The game served as an excellent way of priming our participants, and post-game, they showed excitement in sharing more artifacts from their community. We see this in the significant increase in the new and nuanced artifacts we collected. One of the participants mentioned that ``the game was very enjoyable, and it made us remember all these objects and things that are very implicit to us, and we do not necessarily think about them as being different from us." The richness of artifacts collected varied across participants, which raises an important question of ``whom to consult?" Next, we also observed that states that have large geographic regions, like Karnataka and Uttar Pradesh, have a significant in-state variance in the names of artifacts and sometimes their uses. We also observed that certain social artifacts are similar across neighboring states, and this is in concurrence with our survey questions on ``What other state do you believe is culturally similar to the one that you identify with?"  Unlike human participants, we found that LLMs could lose the logical coreference connections despite having all relevant details in the prompt (refer Fig \ref{fig:prompt}), and adding minimal context like `this \textit{flower} is' helps the LLMs perform better.


\noindent \textbf{Future Work:} This work looks at one dimension of culture - geographic boundaries and takes a step towards introducing how NLP can be made more community-centered. However, there are multiple dimensions, like gender, caste, and race, that shape culture, and LLMs need to be aware of these. Future work should investigate how to evaluate the other dimensions of culture. Given the lack of datasets, especially community-centered datasets, future work should investigate how participatory research can be scaled up to provide more breadth and depth to the datasets created. Creating these datasets would also help preserve knowledge and ensure that LLMs do not lend themselves to cultural erasure. 

\section{Limitations}
Our study is subject to a few important limitations. First, while we cover 18 states, some states and union territories are still missing from the dataset. Second, culture is a highly complex concept shaped by the different identities of individuals, and these intersectionalities of identities lend themselves to the social artifacts considered important by that community. Our work does not systematically recruit from subcultures within each state, making our list of social artifacts incomplete. Third, the language of the survey and the game was English. Since the equivalent name for many objects may not exist in other languages or be unknown to most community members, the language would have limited the responses we got in the survey and impacted the diversity of the participants in both the survey and the game.

\section{Ethical Considerations}
We use the framework by \citet{bender2018data} to discuss the ethical considerations for our work.

\textbf{Institutional Review:} All aspects of this research were reviewed and approved by the Institutional Review Board of a research lab in India.\\

\textbf{Curation Rationale:} To study the cultural familiarity of LLMs, surveys and a Game with a Purpose (GWAP) were conducted. The researchers did not exclude any artifacts given by the survey takers. The vetting by the participants from GWAP ensured that no harmful data made it to the final dataset.\\

\textbf{Language Variety:} The participants for both the survey and a Game with a Purpose (GWAP) culturally identify as Indian. The artifacts are transliterated, and hence, the language would be a mix of (the US variant of English) en-US, (the British variant of English) en-GB, and (the Indian variant of English) en-IN. While transcribing, we transliterated the artifacts and clues to a mix of en-US, en-GB, and en-IN. The researchers were the observers and moderators, ensuring no offensive stereotypes were included in the data.\\

\textbf{Speaker Demographic:} In this version of the study, we do not ask participants to disclose their demographics; this was done to make more participants comfortable in engaging with the questions about their cultural identity and the artifacts they perceive as important.

\textbf{Speech Situation:} The speeches in GWAP were informal, spontaneous, and intended for participants from the same geographical culture.

\section{Bibliographical References}
\bibliographystyle{lrec-coling2024-natbib}
\bibliography{lrec-coling2024-example}

\bibliographystylelanguageresource{lrec-coling2024-natbib}
\bibliographylanguageresource{languageresource}

\setcounter{section}{0}
\renewcommand\thesection{\Alph{section}}
\renewcommand\thesubsection{\thesection.\arabic{subsection}}

\section{Appendix}
\subsection{Survey Questionnaire}
\label{sec:survey_app}
Cultural identity is a way of belonging to a social group with the same intrinsic features and characteristics. Culture is a complex concept, and Individuals consider many factors while constructing their cultural identities: location, sexuality, race, history, nationality, language, religious beliefs, and ethnicity. 

In this study, we ask you to define your cultural identity based on the regions or states in India. Regional identity is a major factor in determining one's cultural identity, so we encourage you to consider this when defining your cultural identity.

Q1.	What is your cultural identity?
\begin{itemize}
    \item Andhra Pradesh
    \item Assam
    \item Bihar
    \item Chhattisgarh
    \item Delhi
    \item Gujarat
    \item Haryana
    \item Jharkhand
    \item Karnataka
    \item Kerala
    \item Madhya Pradesh
    \item Maharashtra
    \item Orissa
    \item Punjab
    \item Rajasthan
    \item Tamil Nadu
    \item Uttar Pradesh
    \item West Bengal
    \item Other
\end{itemize}

Q2. Please list identities that you believe most individuals from your cultural identity would associate with the most or be most familiar with.
\begin{itemize}
    \item Andhra Pradesh
    \item Assam
    \item Bihar
    \item Chhattisgarh
    \item Delhi
    \item Gujarat
    \item Haryana
    \item Jharkhand
    \item Karnataka
    \item Kerala
    \item Madhya Pradesh
    \item Maharashtra
    \item Orissa
    \item Punjab
    \item Rajasthan
    \item Tamil Nadu
    \item Uttar Pradesh
    \item West Bengal
    \item Other
\end{itemize}

Section 2

\textbf{Social Artifacts} As members of a culture, we create objects that help us connect and stay associated with the culture. These objects are known and have significance to most people who consider themselves as a part of that culture and serve as a way of identifying themselves with the culture and the people in that culture. These objects are called social artifacts.

Here are some categories of artifacts that can help guide your thinking.
\begin{enumerate}
    \item  Names of animals like elephants hold importance in Kerela.
    \item  Food and beverages like ghewar hold importance in Rajasthan.
    \item  Clothing like Mysore silk sarees holds importance in Karnataka.
    \item  Home-related ornamentations like Kollam in houses in Tamil Nadu.
    \item Rituals and customs
    \item  Names of handicrafts, dance and music forms, or
    \item Locations like the name of a particular park, shop, monument, beach, etc.
\end{enumerate}

Q3. Write down the 5 `Social Artifacts' that are commonly known, hold relevance to the cultural identity you indicated in question 1, and, to the best of your knowledge, are known to `a significant' number of people who share the same cultural identity.\\

Section 3\\
Q4.	Please list identities that you believe most individuals from your cultural identity would least associate with or be familiar with the least.
\begin{itemize}
    \item Andhra Pradesh
    \item Assam
    \item Bihar
    \item Chhattisgarh
    \item Delhi
    \item Gujarat
    \item Haryana
    \item Jharkhand
    \item Karnataka
    \item Kerala
    \item Madhya Pradesh
    \item Maharashtra
    \item Orissa
    \item Punjab
    \item Rajasthan
    \item Tamil Nadu
    \item Uttar Pradesh
    \item West Bengal
    \item Other
\end{itemize}

Q5. What three other cultural identities are you most familiar with or associate with the most?
\begin{itemize}
    \item Andhra Pradesh
    \item Assam
    \item Bihar
    \item Chhattisgarh
    \item Delhi
    \item Gujarat
    \item Haryana
    \item Jharkhand
    \item Karnataka
    \item Kerala
    \item Madhya Pradesh
    \item Maharashtra
    \item Orissa
    \item Punjab
    \item Rajasthan
    \item Tamil Nadu
    \item Uttar Pradesh
    \item West Bengal
    \item Other
\end{itemize}

\subsection{Survey Data Cleaning}
\label{sec:clean}
The survey response to the question on social artifacts was free-text. Hence, the data was manually cleaned and consolidated to account for differences in spelling and mixed cases for the same artifact. For example, Bandhani and Bandhej refer to the same textile technique and were treated as the same artifact. We did not discard either of the words but treated them as the same artifact.

\subsection{Accuracies at GUESS1 and GUESS2 for Original artifacts}
The accuracy@GUESS1 and accuracy@GUESS2 for original artifacts in DOSA are reported in Table \ref{tab:results_appendix}

\begin{table*}[!htb]
\centering
\begin{tabular}{@{}lllll@{}}
\toprule
\textbf{States} & \multicolumn{4}{c}{\textbf{Models}}                \\ \midrule
\textbf{} & \multicolumn{2}{c}{\textit{Open Source}} & \multicolumn{2}{c}{\textit{Commercial}} \\
          & \textbf{Llama 2}    & \textbf{Falcon}    & \textbf{GPT-4}     & \textbf{Palm 2}    \\ \cmidrule(l){2-5} 
Andhra Pradesh  & 0.07 / 0    & 0.14 / 0 & 0.36 / 0.11 & 0.28 / 0.10 \\
Assam           & 0.18 / 0    & 0 / 0    & 0.36 / 0.14 & 0.27 / 0.14 \\
Bihar           & 0.25 / 0.33 & 0.16 / 0 & 0.42 / 0.14 & 0.50 / 0.50 \\
Chhattisgarh     & 0.11 / 0    & 0 / 0    & 0.22 / 0    & 0.11 / 0    \\
Delhi           & 0.30 / 0    & 0 / 0    & 0.5 / 0     & 0.30 / 0.14 \\
Gujarat         & 0.22 / 0    & 0.05 / 0 & 0.27 / 0.15 & 0.33 / 0    \\
Jharkhand       & 0.33 / 0    & 0.04 / 0 & 0.62 / 0.25 & 0.38 / 0.31 \\
Haryana         & 0.42 / 0    & 0 / 0.08 & 0.58 / 0    & 0.58 / 0.60 \\
Karnataka       & 0.21 / 0    & 0 / 0    & 0.47 / 0    & 0.42 / 0    \\
Kerala          & 0.56 / 0.14 & 0.19 / 0 & 0.63 / 0.33 & 0.63 / 0.16 \\
Maharashtra     & 0.56 / 0.14 & 0.06 / 0 & 0.69 / 0    & 0.44 / 0.44 \\
Odisha          & 0.16 / 0.10 & 0 / 0    & 0.42 / 0.28 & 0.16 / 0    \\
Punjab          & 0.45 / 0.09 & 0.25 / 0 & 0.65 / 0    & 0.55 / 0    \\
Rajasthan          & 0.36 / 0 & 0.18 / 0 & 0.63 / 0    & 0.63 / 0.25    \\
Tamil Nadu      & 0.50 / 0.10 & 0 / 0    & 0.65 /0.14  & 0.45 / 0.27 \\
Telangana       & 0.38 / 0.13 & 0.07 / 0 & 0.46 / 0.14 & 0.38 / 0.13 \\
Uttar Pradesh       & 0.54 / 0.17 & 0 / 0 & 0.38 / 0.25 & 0.54 / 0.33 \\
West Bengal     & 0.25 / 0.20 & 0.05 / 0 & 0.50 / 0.10 & 0.35 / 0.23 \\ \bottomrule
\end{tabular}
\caption{The above table shows guess-wise accuracy on the \textbf{Original Artifacts} in DOSA. The first number shows the \textbf{accuracy@GUESS1}, and the second denotes the \textbf{accuracy@GUESS2}.}
\label{tab:results_appendix}
\end{table*}

\subsection{Benchmarking results for the expanded artifacts in DOSA}
We replicate the benchmarking methodology in Section \ref{sec:benchmark} and calculate the evaluation metrics \ref{sec:eval} for the artifacts collected during the game, i.e., the expanded dataset. For overall accuracy, refer to Table \ref{tab:overall_expanded_artifacts_appendix_table} and Figure \ref{fig:expanded_artifacts_heatmap}. Accuracy@GUESS1 and GUESS2 are in the table
\ref{tab:results_expanded_artfacts_guesswise}

\begin{table*}[!htb]
\centering
\begin{tabular}{@{}lrrrr@{}}
\toprule
\textbf{States} & \multicolumn{4}{c}{\textbf{Models}}         \\ \midrule
 & \multicolumn{2}{c}{\textit{Open Source}}                                   & \multicolumn{2}{c}{\textit{Commercial}}                                  \\
 & \multicolumn{1}{l}{\textbf{Llama 2}} & \multicolumn{1}{l}{\textbf{Falcon}} & \multicolumn{1}{l}{\textbf{GPT-4}} & \multicolumn{1}{l}{\textbf{Palm 2}} \\ \cmidrule(l){2-5} 
Andhra Pradesh  & 0.08 & 0    & \textbf{0.39} & 0.16          \\
Assam           & 0.08 & 0.02 & \textbf{0.09} & 0.02          \\
Bihar           & 0.17 & 0    & \textbf{0.34} & 0.17          \\
Chhattisgarh     & 0    & 0    & 0             & 0             \\
Delhi           & NA    & NA    & NA             & NA             \\
Gujarat         & 0.04 & 0    & 0.07          & \textbf{0.13} \\
Haryana         & 0.12 & 0.07 & 0             & 0             \\
Jharkhand       & 0.34 & 0    & \textbf{0.34} & 0.08          \\
Karnataka       & 0.25 & 0    & \textbf{0.38} & 0.13          \\
Kerala          & 0.31 & 0.16 & \textbf{0.70} & 0.24          \\
Maharashtra     & 0.23 & 0.06 & \textbf{0.62} & 0.45          \\
Odisha          & 0.1  & 0    & \textbf{0.13} & \textbf{0.13} \\
Punjab          & 0.12 & 0    & \textbf{0.48} & 0.28          \\
Rajasthan          & 0.17 & 0.17    & \textbf{0.67} & 0.50         \\
Tamil Nadu      & 0.11 & 0    & \textbf{0.35} & 0.11          \\
Telangana       & 0.12 & 0    & \textbf{0.12} & 0.08          \\
Uttar Pradesh   & 0.06 & 0    & \textbf{0.39} & 0.21          \\
West Bengal     & 0.06 & 0    & \textbf{0.56} & 0.17          \\ \bottomrule
\end{tabular}
\caption{The above table shows the overall accuracy of the \textbf{Expanded Artifacts} in DOSA. (We could not collect any unique expanded artifacts from the participants of Delhi during our interviews.)}
\label{tab:overall_expanded_artifacts_appendix_table}
\end{table*}

\begin{table*}[]
\centering
\begin{tabular}{lllll}
\hline
\textbf{States} & \multicolumn{4}{c}{\textbf{Models}}                \\ \hline
\textbf{}   & \multicolumn{2}{c}{\textit{Open Source}} & \multicolumn{2}{c}{\textit{Commercial}} \\
            & \textbf{Llama 2}    & \textbf{Falcon}    & \textbf{GPT-4}     & \textbf{Palm 2}    \\ \cline{2-5} 
Andhra Pradesh  & 0.08 / 0    & 0 / 0    & 0.15 / 0.27 & 0.15 / 0    \\
Assam           & 0.07 / 0    & 0.02 / 0 & 0.05 / 0.04 & 0.02 / 0    \\
Bihar           & 0.17 / 0    & 0 / 0    & 0.17 / 0.20 & 0.17 / 0    \\
Chattisgarh     & 0 / 0       & 0 / 0    & 0 / 0       & 0 / 0       \\
Delhi           & NA          & NA       & NA          & NA          \\
Gujarat         & 0.03 / 0    & 0 / 0    & 0.06 / 0    & 0.13 / 0    \\
Haryana         & 0.06 / 0.06 & 0 / 0    & 0 / 0       & 0 / 0       \\
Jharkhand       & 0.33 / 0    & 0 / 0    & 0.33 / 0    & 0.08 / 0    \\
Karnataka       & 0.25 / 0    & 0 / 0    & 0.31 / 0.09 & 0.13 / 0    \\
Kerala          & 0.15 / 0.18 & 0.15 / 0 & 0.61 / 0.20 & 0.23 / 0    \\
Maharashtra & 0.22 / 0            & 0.05 / 0           & 0.44 / 0.30        & 0.38 / 0.09        \\
Odisha          & 0.09 / 0    & 0 / 0    & 0.13 / 0    & 0.13 / 0    \\
Punjab          & 0.12 / 0    & 0 / 0    & 0.40 / 0.13 & 0.28 / 0    \\
Rajasthan          & 0.17 / 0    & 0.17 / 0    & 0 / 0.67 & 0.17 / 0.40    \\
Tamil Nadu      & 0.07 / 0.03 & 0 / 0    & 0.20 / 0.17 & 0.10 / 0    \\
Telangana       & 0.11 / 0    & 0 / 0    & 0.11 / 0    & 0.07 / 0    \\
Uttar Pradesh   & 0.05 / 0    & 0 / 0    & 0.38 / 0    & 0.11 / 0.07 \\
West Bengal     & 0.05 / 0    & 0 / 0    & 0.38 / 0.27 & 0.16 / 0    \\ \hline
\end{tabular}
\caption{The above table shows guess-wise accuracy on the \textbf{Expanded Artifacts}. The first number shows the \textbf{accuracy@GUESS1}, and the second denotes the \textbf{accuracy@GUESS2}.}
\label{tab:results_expanded_artfacts_guesswise}
\end{table*}

\begin{table*}[]
\begin{tabular}{@{}c|c|l@{}}
\toprule
\textbf{State} &
  \textbf{Artifact} &
  \multicolumn{1}{c}{\textbf{Clues}} \\ \midrule
\multirow{3.25}{*}{Punjab} &
  Lohri &
  \begin{tabular}[c]{@{}l@{}}this is a Punjabi festival usually celebrated in winters\\  people make bonfires and kites are flown during the day\\  this festival is also known as festival of kites\\  people usually eat peanuts, rewaris etc. during this festival\end{tabular} \\ \cmidrule(l){2-3} 
 &
  sarson ka saag &
  \begin{tabular}[c]{@{}l@{}}this is a food item that is usually had in winters\\ It is made using the leaves of the mustard plant\end{tabular} \\ \midrule
\multirow{3.25}{*}{Maharashtra} &
  Lavani &
  \begin{tabular}[c]{@{}l@{}}It is an old form of dance\\ This form of dance is usually practiced when the villagers \\ are done harvesting their crops\end{tabular} \\ \cmidrule(l){2-3} 
 &
  Tandalachi bhakri &
  \begin{tabular}[c]{@{}l@{}}this is an alternative to roti\\ it is made of rice\\ it is white in color\end{tabular} \\ \midrule
\multirow{3.25}{*}{Assam} &
  Jappi &
  \begin{tabular}[c]{@{}l@{}}A headgear usually worn by farmers, fishermen, and tea garden workers.\\ These days, it has been commodified for gifting.\\ It is made of bamboo straws, and the ones that are souvenirs \\ sometimes have designs made of velvet clothes.\end{tabular} \\ \cmidrule(l){2-3} 
 &
  Gamusa &
  \begin{tabular}[c]{@{}l@{}}A traditional garment which is usually used to wipe oneself.\\ Traditional ones are white and red in color.\\ The patterns or designs on it are used to distinguish the use of this garment.\end{tabular} \\ \midrule
\multirow{3.25}{*}{Tamil Nadu} &
  Veshti &
  \begin{tabular}[c]{@{}l@{}}It is usually a part of men's attire\\ It is a kind of Long cloth\end{tabular} \\ \cmidrule(l){2-3} 
 &
  kolam &
  \begin{tabular}[c]{@{}l@{}}It's found in a Tamil house\\ Outside the door area\\ Made out of white powder\end{tabular} \\ \bottomrule
\end{tabular}
\caption{The above table shows a few of the artifacts and their corresponding clues collected during the participatory research}
\label{tab:artifacts_table}
\end{table*}


\end{document}